# HARMONI at ELT: designing a laser guide star wavefront sensors for the ELT

Anne Costille[*a], Anne Bonnefoi[a], Edgard Renault[a], William Ceria[a], Kjetil Dohlen[a], Benoit Neichel[a], Zoltan Hubert[b], Jean-Jacques Correia, Thibaut Moulin, Saul Menendez Mendoza, Thierry Fusco, Pascal Vola, Felipe Pedreros, Pierre Jouve, Kacem El Hadi, Fraser Clarke[e], Hermine Schnetler[f], Dave Melotte[f], Niranjan Thatte[e]

[a] Aix Marseille Univ, CNRS, CNES, LAM, Marseille, France
[b] Durham University, Centre for Advanced Instrumentation, Durham DH1 3LE, United Kingdom
[c] Instituto de Astrofísica de Canarias (IAC) C. Vía Láctea, 1, 38205 La Laguna, Santa Cruz de Tenerife, Spain
[d] ONERA, B.P.72, 92322 Chatillon, France
[e] Department of Physic, University of Oxford, Oxford OX1 3RH, United Kingdom
[f] UKATC, Royal Observatory Edinburgh Blackford Hill, Edinburgh EH9 3HJ, United Kingdom

**ABSTRACT**

HARMONI is the first light visible and near-IR integral field spectrograph for the ELT covering a large spectral range from 450nm to 2450nm with resolving powers from 3500 to 18000 and spatial sampling from 60mas to 4mas. It can operate in two Adaptive Optics modes - SCAO and LTAO - or with no AO. The project is preparing for Final Design Reviews. The laser Tomographic AO (LTAO) system provides AO correction with very high sky-coverage thanks to two systems: the Laser Guide Star Sensors (LGSS) and the Natural Guide Star Sensors (NGSS). LGSS is dedicated to the analysis of the wavefront coming from 6 laser guide stars created by the ELT. It is made of 6 independent wavefront sensor (WFS) modules mounted on a rotator of 600mm diameter to stabilise the pupil onto the microlens array in front of the detector. The optical design accepts elongated spots of up to 16 arcsec with no truncation using a CMOS detector from SONY. We will present the final optical and mechanical design of the LGSS based on freeform lenses to minimize the numbers of optical components and to accommodate for the diversity of sodium layer configurations. We will focus on rotator design, illustrating how we will move 1 tons with 90" accuracy in restrictive environment. Finally, we will present the strategy to verify the system in HARMONI context. The main challenge for the verification being how to test an AO system without access to the deformable mirror, part of the ELT.

**Keywords:** ELT Harmoni, Adaptive Optics, Opto-Mechanics, LTAO, laser guide star, AIT, wavefront sensors

## 1. INTRODUCTION

HARMONI [1] is a visible and near-infrared integral field spectrograph providing the ELT's core spectroscopic capability. It will exploit the ELT's scientific niche in its early years, starting at first light. To get the full sensitivity and spatial resolution gain, HARMONI will work at diffraction limited scales. This will be possible thanks to two complementary adaptive optics systems [2], fully integrated within HARMONI. The first is a Single Conjugate AO (SCAO) system offering high performance for a limited sky coverage. The second is a Laser Tomographic AO (LTAO) system, providing AO correction with a very high sky-coverage. Both AO modes for HARMONI have gone through Preliminary Design Review at the end of 2017, and enter their Final Design phase starting in 2018 ending in 2023.

---

[*] E-mail: anne.costille@lam.fr

While the deformable mirror performing real-time correction of the atmospheric disturbances is located within the telescope itself, instruments are in charge of providing the wavefront measurements controlling this correction. To this end, HARMONI contains a suite of state-of-the-art and innovative wavefront sensor systems. These are distributed within the instrument according to their various functions. The Laser Guide Star Sensors (LGSS) are located at the entrance of the instrument and fed by a dichroic beam splitter reflecting the sodium line and holds by the optical relay at the entrance of the instrument. The various natural guide star sensors (NGSS) are located close to the science focal plane. LTAO requires natural guide stars for fast sensing of tip-tilt and focus and for disentangling the effects of slowly varying non-common path errors.

In this paper, we present the current design of the LGSS system in particular the optical design, described in more details in [3], and the mechanical design, with particular attention given to some of the more original aspects of the designs. We will also present the assembly, Integration and test approach chosen for the system.

## 2. LASER GUIDE STAR SENSORS OVERVIEW

### 2.1. HARMONI instrument

The HARMONI instrument (Figure 1) is built up of four main hardware systems. The integral field spectrograph (IFS) [4] consists of a 4m high and 3.25m diameter cylindrical cryostat vessel sitting on a rotator, housing scale-changing optics, image slicer units and spectrograph modules. The calibration and relay system (CARS) consists of a relay system (FPRS) in the form of a cooled Offner relay producing a vertical optical axis and a focal plane at the entrance of the IFS, a calibration unit, and the instrument structure (ISS) allowing to present these elements to the telescope optical axis, 6m above the Nasmyth platform. The two remaining systems are the wavefront sensor systems, one working on artificial laser guide stars (LGSS), and the other working on natural guide stars (NGSS).

To implement the LTAO operational mode, HARMONI instrument uses the laser guide stars sensors provided by the LGSS together with the appropriate natural guide start sensor (LOWF subsystem part of NGSS (Natural Guide Star Sensors)), the ELT corrective optical components (tip-tilt mirror (M4) and deformable mirror (M5)) together with the Adaptive Optics Control System (AOCS) to configure the instrument. Once configure the virtual instantaneous LTAO system is used to correct the incoming wavefront, stabilize the science focal plane and communicate with the ELT CCS to track the science object of interest across the sky

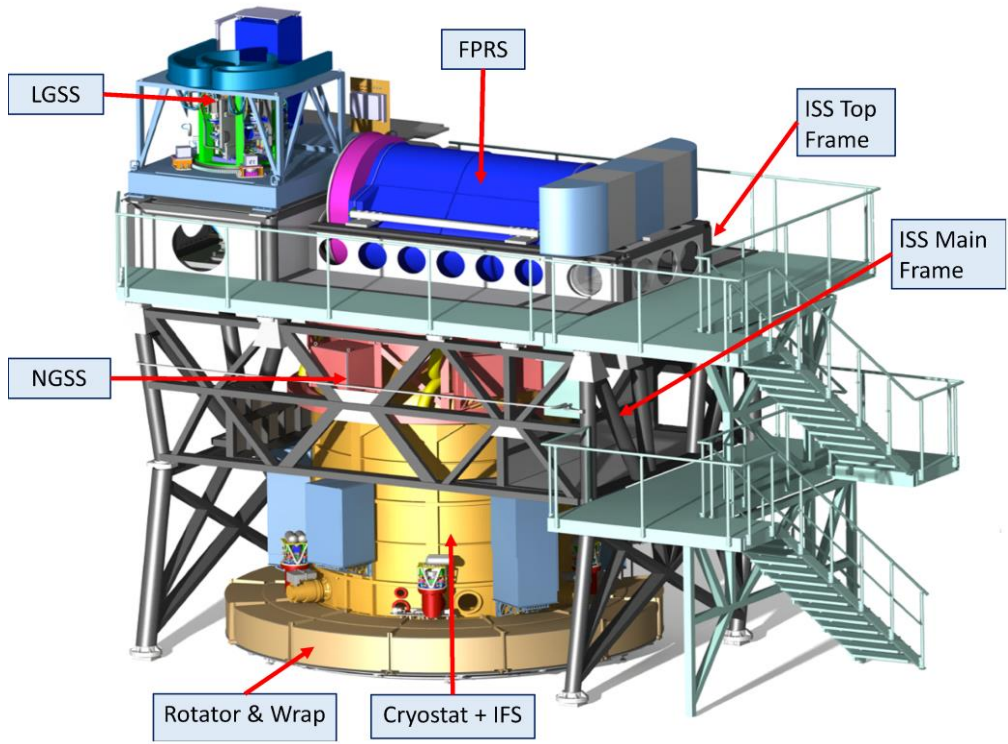

Figure 1. 3D rendering of the HARMONI instrument indicating major systems and components.

**2.2. Laser guide-star sensors design choice**

The six ELT laser guide stars are analysed using the LGSS system. Located at the entrance of HARMONI, it receives the laser light reflected off a large dichroic beam splitter located 1m downstream of the telescope focus. A fold mirror sends the 6 laser beams vertically upwards where they are individually intercepted by the six LGS Wavefront sensor Modules (LWM). These are mounted in a rotating core structure, controlled such as to maintain the pupil stabilized within the wavefront sensors.

**2.2.1.   LGS asterism**

During the PDR phasis [5], the LGSS proposed a design based on a variable asterism of the LGS constellation. The choice was motivated at that time by the goal to have the best performance in function of the Zenith angle. Following the HARMONI PDR and the formal acceptance by ESO of the LTAO mode, the design choice of the whole LTAO mode has been reviewed and improved. The main change was to go to a fixed LGS constellation. The trade-off between the LGS constellations (optimal or fixed), which comes to a trade-off between optimizing the performance on-axis (science), and off-axis (NGS direction) shows that there is a small gain in using the fixed constellation as an open loop deformable mirror is present in the LOWFS arm [6]. This is illustrated by Figure 2. The design presented in this document is then based on a fixed asterism.

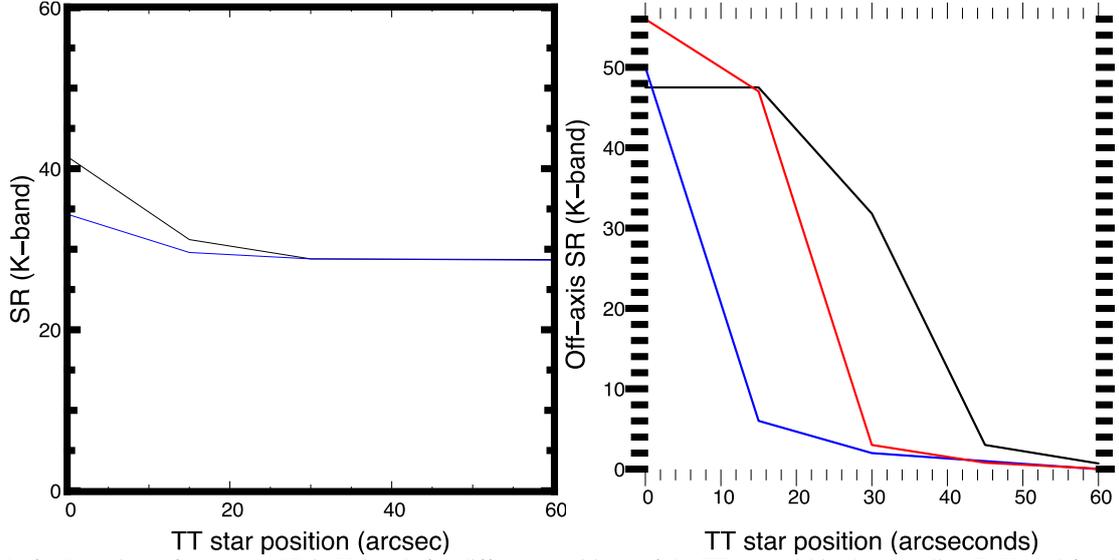

Figure 2: Left: On-axis performance (SR in K-band) for different positions of the TT star within the patroling FoV, and for the optimal LGS constellation (black) and the fixed LGS constellation (blue). Right: Off-axis performance (SR in K-band), at the NGS star position, for no OLDM (blue), perfect OLDM with fixed LGS constellation (black), perfect OLDM with optimal LGS constellation (red).

The choice of a fixed asterism requires the definition of the value of the asterism chosen for the LGSS constellation. This value is a compromise between optimisation of the LTAO performance and the feasibility of the opto-mechanical design. The main difficulty for LGSS design is to be able to separate the six LGS beams at the entrance of the WFS benches and to stay in the allocated volume of Harmoni. In terms of LTAO performance, analysis indicate that the asterism shall have a value between 30" and 35" to have the best trade-off between on-axis and off-axis tomographic optimization, with a preference to have small asterism. The opto-mechanical constraints on LGSS have driven the choice of the final value of the asterism. In particular, the distance between the LGS focus and the first lenses at Zenith allows to separate properly the beams at the entrance of each bench. Therefore, in Harmoni case, the asterism radius has been fixed to 34".

### 2.2.2. WFS camera choice

A key driver of LTAO performance is the LGS WFS camera and associated detector. It is then needed to use for the LGSS system the camera that is best developed and optimised for LGS sensing. A major difficulty with the sodium laser guide star for a 39 m telescope is the spot elongation. The laser "star" formed in the ~20km thick atmospheric sodium layer some 90km above the telescope actually is a luminous cylinder. Launched from the edge of the pupil, it appears as a streak of angular width of 20" when observed from the opposite edge of the pupil.

Ideally, one would use a large enough detector array, providing a FOV per sub-aperture larger than 20" in order to fit the entire streak. With a pixel scale of 1"/pixel, dictated by the typical angular width of the laser star, this would require 20N square pixels, where N is the number of sub-apertures across the WFS. Since we consider N=68 for the LGS-WFS, this would require an array of at least 1360x1360 pixels. Two detectors have been studied for the LGSS design: the LVSM detector provide in a LISA camera from ESO [7] and SONY detector provided in a CBLUE1 camera by FLI [8]. We have compared the two detector in term of performance and the project has finally chosen the SONY detector after several studies as it is offering the best performance for the project [9, 10]. The summary of the key requirements can be found in Table 1. The key requirements are: to have a low noise detector, with as much pixel as possible to provide a large FoV per subaperture to limit the truncation of the elongated LGS spot and to have a global shutter.

Table 1. WFS Detector requirements

|  | HARMONI requirements | LVSM detector | SONY detector |
|---|---|---|---|
| Detector size | > 800 x 800 pixels<br>Goal : 2000 x 2000 pixels | 800 x 800 pixels | 1608 x 1136 pixels |
| Pixel size | Between 10 and 40 µm | 24µm | 9µm |
| Reading mode | Preferred global shutter | Rolling shutter | Global shutter |
| Quantum Efficiency @ 589 | Baseline > 70%, Goal > | > 85% (camera | QE x FillFactor = 73% |
| nm | 95% | specification) | (measured value) |
| RON | < 3e- (Goal 1e-) | Measured: ~3.1e- | Measured: 2.77 +/-0.0773e- |
| Frame rate (Baseline) | From 100Hz to 500Hz with selectable frequencies every 100Hz. Goal is up to 1000Hz | 700Hz, full frame | 480Hz for 1100x1100 pixels in 12bits. 660Hz for 1100x1100 pixels in 9bits |
| Dark current | < 50e-/s/pix | ~90e- (not cooled system) | 46e-/s/pix |
| Cosmetics | Baseline = Less than 1 bad / dead pixel (defined as a deviation of more than 20 % of a nominal response curve) per bunch of 10x10 adjacent pixels. | < 0.1% (camera specification) | 0 dead pixels. 545 hot pixels on the whole image. 53 pixels with more than 30% deviation == 0,034 hot pixel on each 10x10 pixels zones. |
| Angle of acceptance | | +/-30° | Specified at +/-10° Measured at +/-20° |

### 2.2.3. Microlens design

The LGS-WFS detector is based on a SONY detector based on CMOS technology of 1608 x 1136 pixels. One main difficulty of the SONY detector is the small size of the pixel, 9μm pitch, and the associated acceptance angle of the pixel. The acceptance angle is linked to the maximum focal ratio that can be accepted by the detector. In the case of the LGSS, the focal ratio will be close to 3 leading to an acceptance angle on the microlens array around 16°. Figure 3 presents the acceptance angle of the SONY detector in horizontal and vertical directions provided by the manufacturer. These angles have been experimentally measured at LAM on a SONY detector and have shown very similar results [9]. We confirm that the acceptance angle is not symmetric in horizontal and vertical direction. If the maximum angle on the detector is around 16°, it corresponds to the worst case in a collecting light of 40% which is not good. A lot of light will be lost by the detector and then it will affect the performance.

According to LTAO analysis, impact on LTAO performance is limited if the light collecting efficiency is larger than 80%. It is then necessary to have acceptance angle smaller than +/-10° in the worst case to limit effect on LTAO performance with a SONY detector. To decrease the maximum angle arriving on the detector from LGSS, we propose to use a double microlens concept in front of the detector. Figure 4 shows the ray tracing for classical microlens array and double microlens array concept. This concept has the advantage to create a telecentric image, and when using with an optical relay between the microlens array and the detector, to create a pupil in the relay. In this case, the image is telecentric and then the input angle on the detector reduced by a factor 2 leading an angle of 8°.

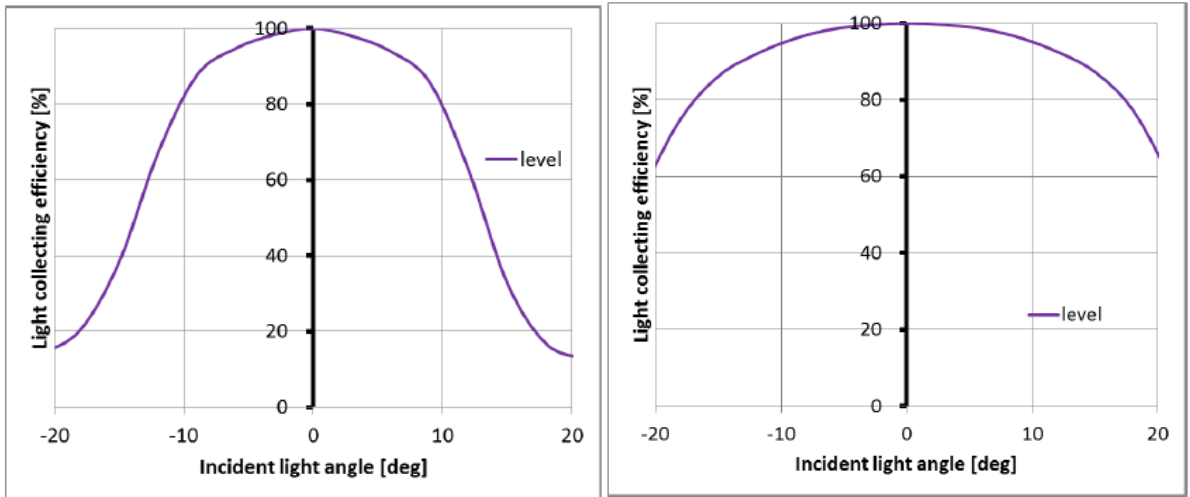

Figure 3. Acceptance angle of the SONY detector from SONY datasheet.

We are then in the acceptable range of angle of acceptance for the detector with a limited impact on LTAO performance. It has then been decided to use a double microlens array for Harmoni as double microlens array can be manufactured such as in BIGRE spectrograph for SPHERE IFS thanks to AMUS company.

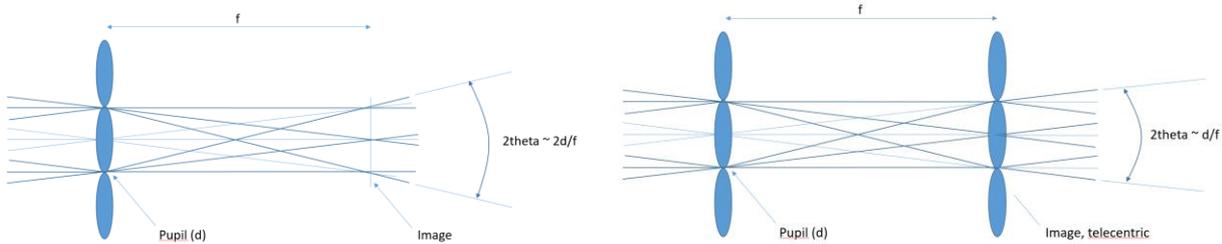

Figure 4: Acceptance angle: Left: classical microlens array in front of a detector. Right: double microlens array concept with telecentric image.

## 3. LGSS OPTO-MECHANICAL DESIGN

### 3.1. System overview
Final design Review of the LGSS design is foreseen for June 2023 so the design of the system is now mature. We propose in the following sections a detailed overview of the choice made for the design. The LGSS is made in 3 main sub-systems (see Figure 5):
- The LGSS support frame contents all the mechanical elements that compose the structure of the LGSS. It includes a rotator to allow the stabilization of the pupil onto the microlens array and a cable wrap to allow the cables to move properly. It holds the LGSS cabinets and the service routing that provides power, signal and glycol to the different parts of LGSS.
- The LGS fold mirror reflects the light of the laser vertically and that is directly installed onto the ISS of HARMONI.
- The LGSS rotating sub-system holds the 6 optical benches of the LGSS. This is the opto-mechanical core of the system where the WFS are installed and analyze the light coming from the LGS. The LGS rotating sub-system is also structure in three main modules
    o The LGS core structure made in stainless steel and aligned to the rotator axis. It holds 6 WFS benches and the cable wrap
    o The LGS WFS modules (LWM) hold the optical component to focus the light and conjugate the pupil plane on to the detector array
    o The LGS Detector Modules (LDM) made of a microlens array, a focal plane relay and a detector.

The mass of LGSS is limited to 2.5T (excluding cabinet mass).

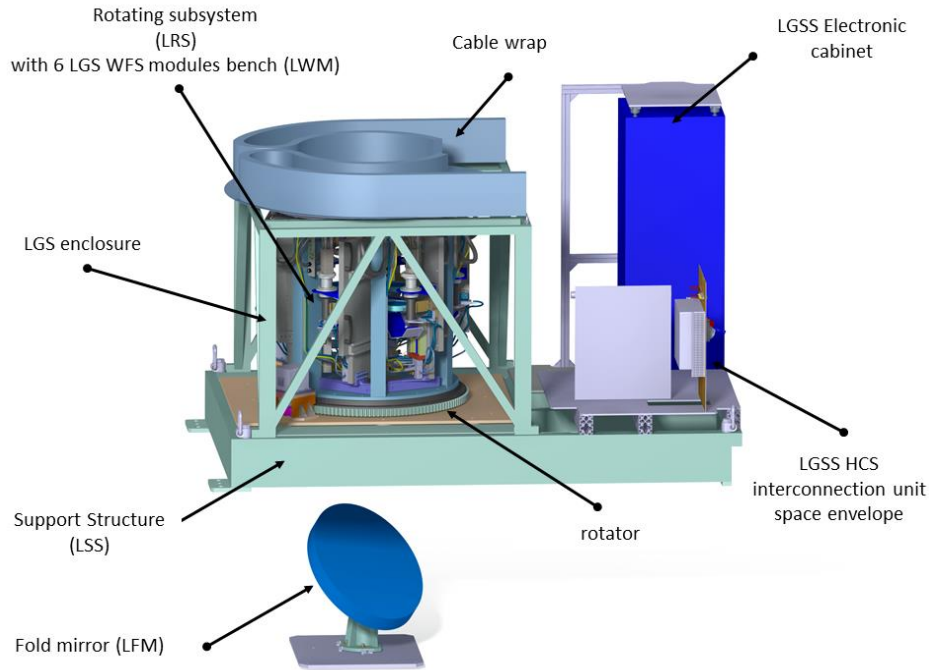

Figure 5: LGSS system overview

### 3.2. Optical design

The most critical part of the optical design for LGSS are the pupil imaging lenses at the entrance. They need to produce well-corrected images at vastly different object distances (the laser focus moves by several meters when going from Zenith to 30 degrees from the horizon). An appropriate design has been found using four free-form lenses as shown in **Erreur ! Source du renvoi introuvable.**, staying within the required wavefront error budget of 120nm rms and 0.1% pupil distortion. The design and its performance analysis is detailed in [3]. We only give here an overview and its main characteristics. The LGSS design needs to accommodate for:

- Sodium layer medium altitude above telescope: 87.8km +/- 2km at Za = 0°
- Sodium layer thickness: +/-11km
- Pupil conjugation with microlens array: M1 mirror with entrance pupil size: 38542mm
- Exit pupil distance: 37849mm
- Telescope Zenith angle: 0 to 60°
- Compatibility with a fixed asterism of 34'' radius
- Provide a field of view non vignetted ≤ 16''. A field stop shall be implemented
- Providing focus correction for layers ≥ 84.8km and ≤ 92km at Za = 0°
    - Layer < 84.8km will be decentered on the WFS but not vignetted
    - Layers >184km will be decentered on the WFS but not vignette
    - The rest of the layer altitude will be truncated
- Providing a pupil stabilization with an accuracy of 0.03 sub-apertures
- Providing a bandapss filter of 589nm +/-5nm
- Capacity to handle the non-common path errors with respect to science path
- Global transmission ≥ 45% (detector quantum efficiencly included)

According to the requirements described below, an optical design has been proposed for the LGSS system [3]. Light of the LGS is received in the horizontal plane coming from the dichroic beam splitter which is part of the CARS system. A large folding flat which is part of LGSS redirects the beam vertically into the rotating structure holding the six LGS WFS modules (LWM). The optical design of each LWM is composed of the following elements:

- Field optics, placed in the space between the infinite conjugate focus and the LGS focus. Creates a pupil image and reduce the focal distance in order to fit with the volume allocation Freeform lenses are needed to achieve the requested image quality.

- Folding mirrors to fold the optical beam and fit the optical design in the LWM dimensions
- Camera optics to create a telecentric beam with a unique focal ratio. Two freeform lenses are implemented to control pupil distortion
- Focussing trombone with a mechanism to compensate the focus difference between zenith and 60° pointing. The maximal shift (between these two pointing angles) is 97mm.
- A Pupil Control lens to align accurately the pupil onto the microlens array
- A square Field Stop defines the FoV (16" on sky) and orientes according to the WFS camera orientation
- Pupil Objective Lenses to create a pupil of 22.58 mm diameter at the microlens array.
- Detector module made of the microlens array, a relay optics and a SONY detector. Raytracing diagrams

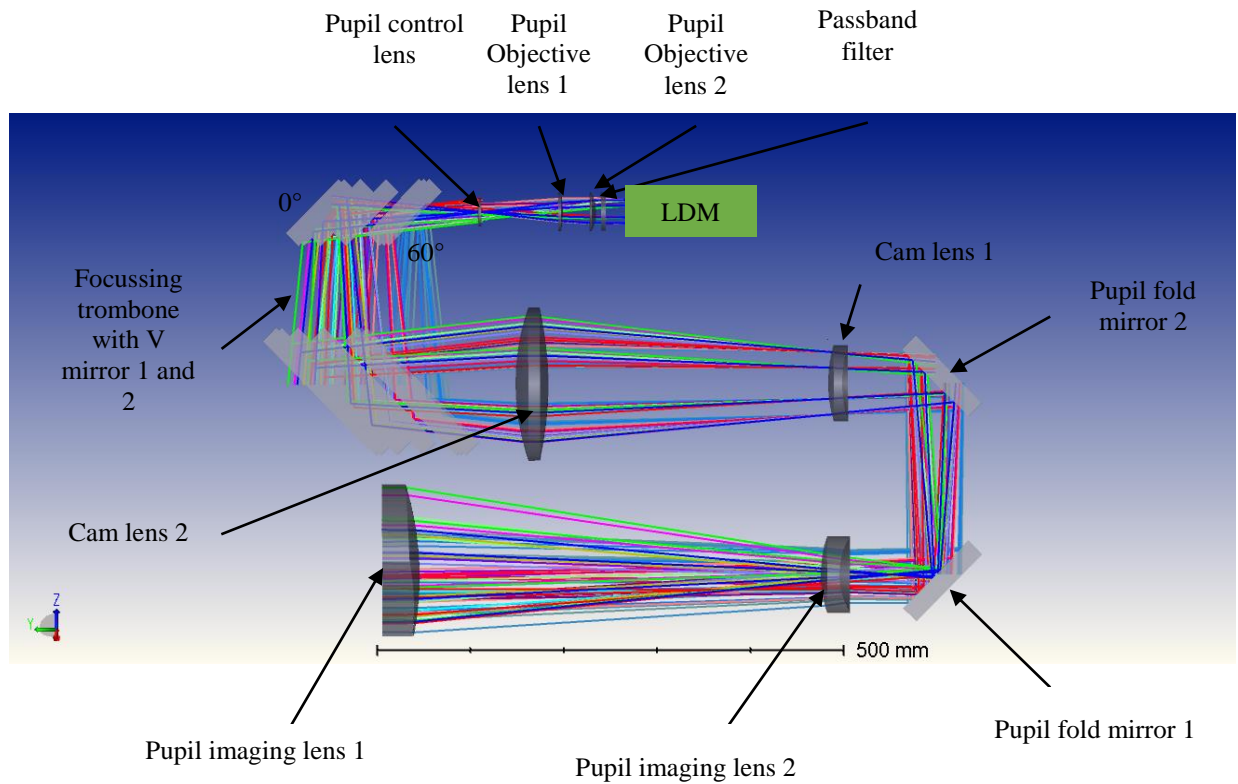

Figure 6: Ray tracing diagram of the LWM up to the microlens array. The fold mirror is not shown.

### 3.3. LGSS detector module description

The LGSS Detector Module (LDM) contains a microlens array, relay optics, and a camera package. The detector module is developed by IPAG, with a joint effort and analysis with MORFEO project. The baseline for the LGSS design is to use a SONY detector, delivered in a C-BLUE 1 camera. The current design of LGSS provides a detector module directly installed on the optical bench modules, which is different from the PDR design where the detector module was separated from the benches. The main characteristics of the detector module are:

- 68x68 sub-aperture in the pupil
- Minimum useful FoV per sub-aperture: 16''
- PSF sampling on the detector > 1.1''/pixel
- Double-sided microlens array to limit the angle of acceptance on the detector

- Optical relay: Due to the wide field required for LGS wavefront sensors, the microlens focal length is very short. To avoid implementing a microlens inside the detector chip, it is therefore mandatory to implement a relay system to reimage the Shack-Hartmann spots onto the detector. This problem is common with other ELT instruments, and an agreement is made with the MORFEO team to develop common detector system module: microlens, relay, camera. Figure 7 shows ray paths traced through the design of a relay with a C-BLUE 1 camera. The length of this system is about 150 mm, from lenslet array to detector.

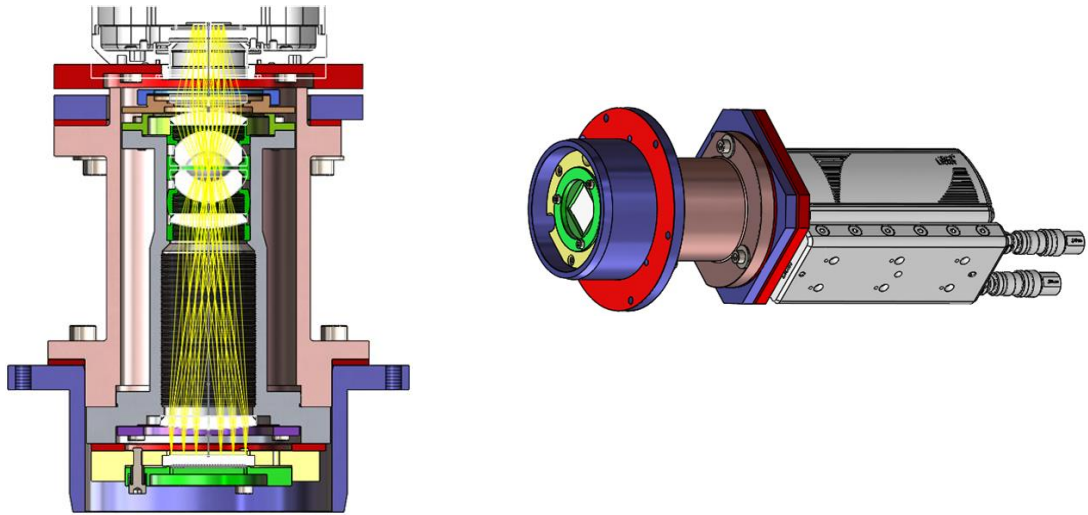

Figure 7. Left: ray tracing and opto-mechanical concept of the LDM. Right: global CAD view of the LDM with the CBLUE1 camera.

**3.4. LGSS mechanical implementation**

The design of the LGSS has a lot evolved since PDR and several modifications have been proposed to the design to make it more robust and easier to align. The main changes are:

- The separation of the LGS Fold Mirror from the LGSS main structure. The FM is put directly in interface with the HARMONI ISS structure. It provides a better and more stable interface, it allows to make the LGSS alignment easier by decoupling the alignment of the rotating structure from the FM. It also facilitates the integration and handling of the system.
- The LGS detector module is no longer set on the bottom of the core structure, but directly installed onto the WFS bench.
- The rotator is now located at the bottom of the core structure directly on the structure. It is possible due to the change of the optical path design. This allows to lighten the global LGSS structure and to have the mass of the core structure supported by the rotator, and not "hanged" to it.

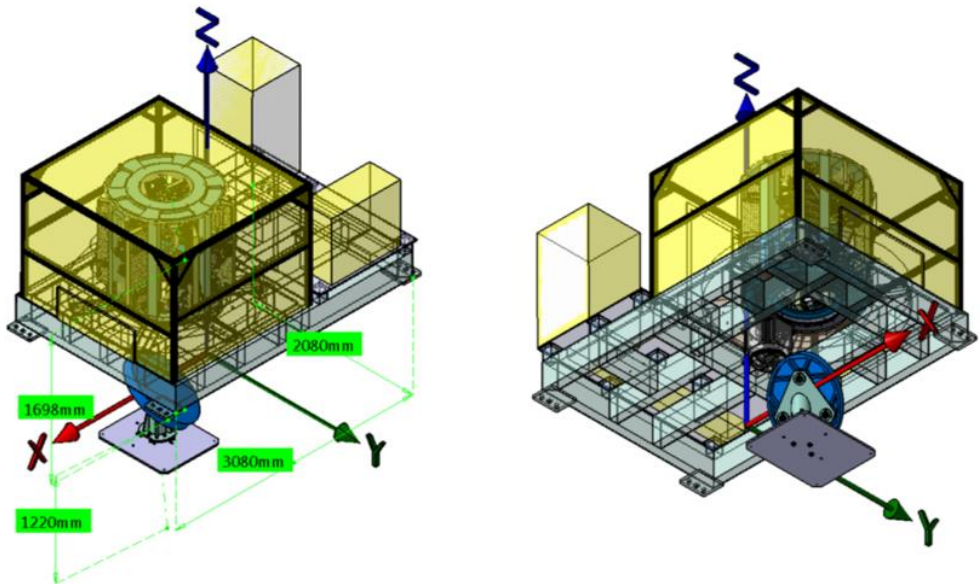

Figure 8: LGSS global layout viewed from top (left) and from bottom (right). Cable wrap is not shown on this picture

### 3.4.1. LGSS Fold Mirror

The fold mirror is fixed on the HARMONI ISS platform. The mirror is elliptical, 860mm long, 600mm wide. The support will be as stiff as possible to limit Fold mirror displacement. The baseline for the LGSS fold mirror development is to sub-contract the detailed design of the fold mirror, in particular how the mirror will be maintained on its holder. Optical manufacturers are skilled to develop such mirrors and to take into account our constraints in term of tenue to earthquake, environmental conditions and also to avoid deformation of the optical surface. The mirror is quite large and then could be quite heavy if manufactured in full part. Our current analysis proposes a design of a lighten mirror with a mass lower or equal to 70kg. Different manufacturers (Thales SESO, Winlight Optics) have confirmed the feasibility to manufacture a fold mirror with a mass lower than 70kg, with a surface flatness quality better than 100nm rms on the reflected beam at 45° for any beam of 250mm diameter over the optical design. Fold Mirror support. Our proposed strategy for holding the mirror is to glue three Invar® pads on the back of mirror (Figure 9). The Invar is the metallic material which CTE is the closest from the one of Zerodur®. Each pad includes a blade which normal is directed toward the axis of symmetry of the mirror. The three pads are then screwed to a square made of steel (baseline) or of an aluminium alloy. Development of the design of the pads will be done together with the mirror manufacturer.

The base of this square is finally used for providing an adjustment along 6 DOF with a set of shims, and the screwing on the ISS fold mirror interface.

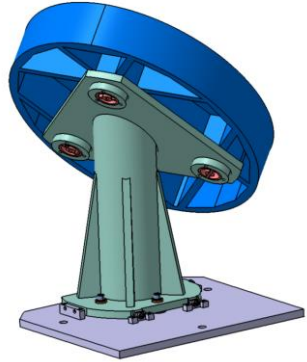

Figure 9: Preliminary design of the fold mirror.

**3.4.2. LGSS support structure**

The LGSS support structure is one of the main LGSS sub-systems. It carries the rotator and the core structure, the LGS enclosure and the interfaces for the elements of HARMONI Control System (HCS) carried by LGSS. Figure 10 shows the current design of the LGS support frame. The four bases of the feet are linked to HARMONI structure via four adjustable pads. They support 6 tons and allow a vertical adjustment of +/- 5 mm (tbc) and accept an orientation defect of 3 degrees. If the LGSS is dismounted, three mechanical stops (two along X and one along Y) are not unscrewed: they constitute the reference for later repositioning. The design of the bases if common to other HARMONI interfaces.

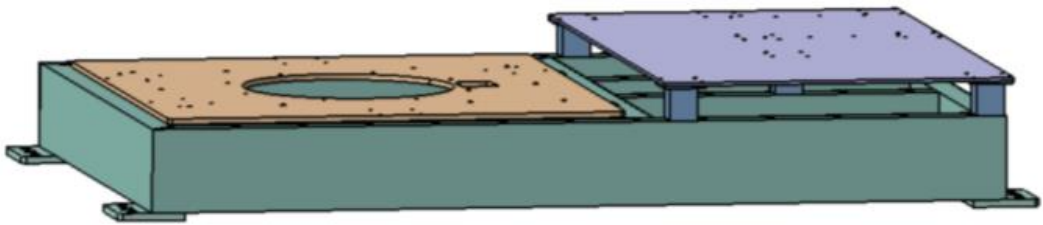

Figure 10: Global layout of LGSS structure (green) supporting the cabinets (blue) and the derotator (tan)

The LGSS core structure and the rotator will be placed into an enclosure to provide a protection for dust and light coming from the Nasmyth platform. This will not be a complete hermetic structure but the contamination level will be limited. Access door will be implemented to be able to remove the LWM bench from the core structure.

The LGS support frame carries also some elements from HCS to support LGSS operation. In particular, LGSS owns a cabinet which contains of the controllers for the mechanisms, the power and trigger signal for the WFS cameras, the calibration source module. Cables are distributed within the system and access the main benches with the cable wrap. An interconnection unit is provided to safely disconnect LGSS from the rest of HARMONI in case of maintenance activity. This unit provide power, network and glycol distribution for the whole LGSS. Finite element analysis are on going to characterize the behaviour of the system. In particular the eigen mode are studied to verify that they are higher than 16Hz. For the moment, the first mode is around 12.5Hz as shown in Figure 11 but this is not a concern for the design.

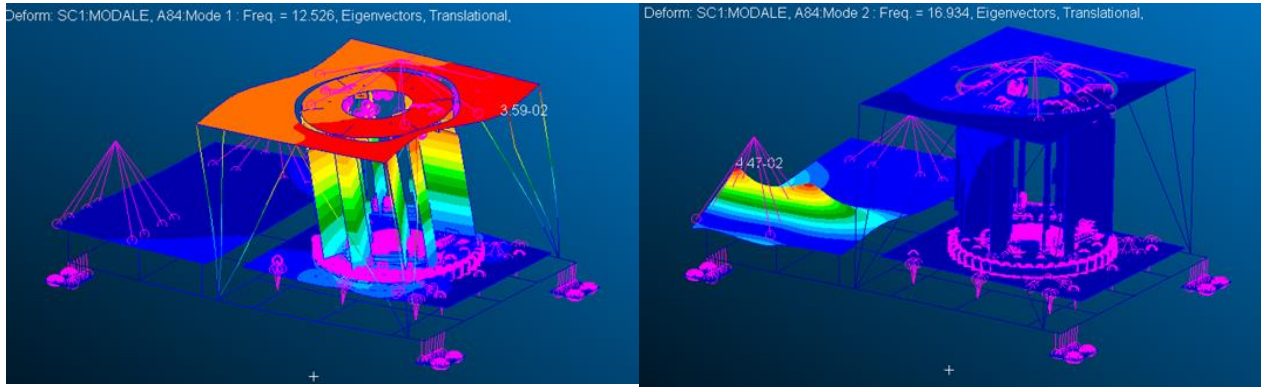

Figure 11: FE mode analysis on LGSS. Left : first mode at 12.5Hz, Right: second mode 17Hz

### 3.4.3. LGSS rotator

LGS rotator is one of the critical items for LGSS. Its design is under study by LGSS mechanical team together with HARMONI control system team to define the control solution. The rotator synchronizes the movement of the wave-front sensors to the apparent movement of the M4 deformable mirror in the telescope. In principle this movement follows the altitude movement of the telescope, governed by sidereal tracking laws. However, with a requirement of 0.03 sub-aperture stability of the microlens array with respect to M4 actuators, and accounting for apparent M4 movements due to telescope, platform, and instrument flexures, following the classical pupil tracking law will not be sufficient. The real time controller will be able to produce an error signal for pupil rotation which must be fed back to the rotator motion controller. The main requirements are:

- Range:
    o An operational range of 90° following possible variation of the telescope altitude;
    o A maintenance range of at least 340°. This is to allow maintenance on each LWM and access to them.
- Precision / needed accuracy: +/- 90'' (LGSS rotator shall align the pupil in rotation better than +/-0.03 sub-apertures)
- Minimum incremental movement (resolution): +/-30". We allocate 1/5th of the needed accuracy in as error due the resolution of the rotator mechanism
- Wobble: < +/-50'' on any course of 90°
- Run-out: < +/-100µm on any course of 90°
- Angular speed:
    o In operation: telescope altitude variation from 0.0001arcmin/s to 0.25arcmin/s.
    o In maintenance: > 1°/s

The proposed design is based on custom bearing from Thyssengroup. 2 motors with gear boxes (from Kollmorgen) are implemented and will work in opposition to compensate the backlash error. The rotator is based on the bottom of the LGSS core structure and has to move more than 800kg taking into account the cable wrap. We have added an Heidenheim encoder and limit switches to control the rotation and protect the cable wrap to cable pulling.

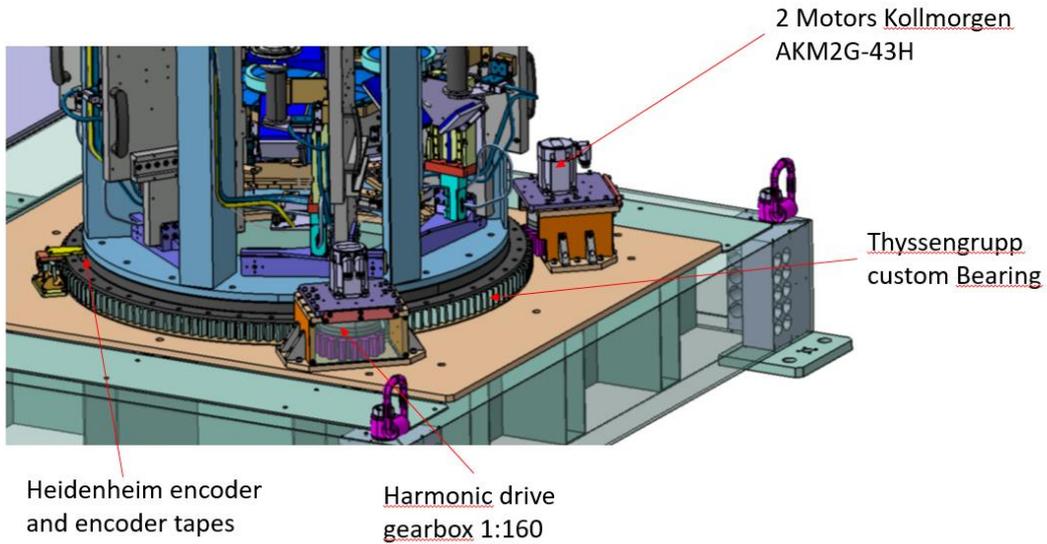

Figure 12: Overview of the LGSS rotator design and the main parts.

#### 3.4.4. LGSS rotating sub-system
**Erreur ! Source du renvoi introuvable.** shows the LGS rotating sub-system. This assembly gathers the rotating components of the system.

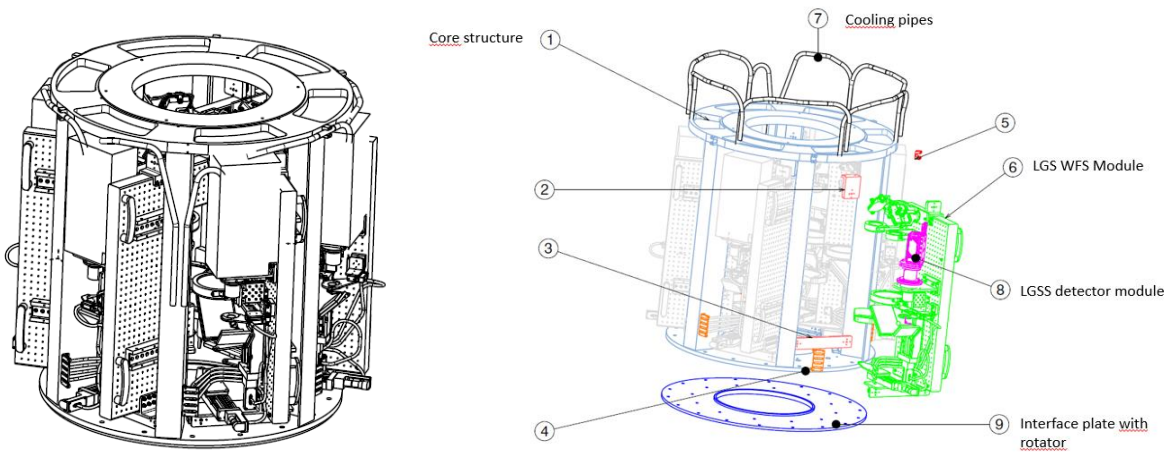

Figure 13: The opto-mechanical assembly of the LGS rotating sub-system

The *Core Structure* interfaces the rotator and supports the six optical benches and the WFS cameras. It is also used to support the cables to feed the mechanisms and the cameras. Its geometry is more or less directed by the six LGS optical beams. The material selected for the design is stainless steel to avoid issue due to thermal expansions between the rotator and the benches and to have a first natural frequency as high as possible. **Erreur ! Source du renvoi introuvable.** shows the current design of the core structure.

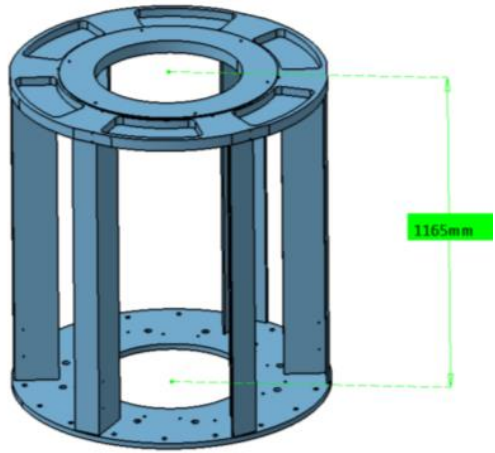

Figure 14: Core structure layout in steel.

Figure 15 shows the CAD view of the LWM including the LDM. The optical bench is 895 mm long and 280 mm wide. There are six of them identical in the system, paired with the 6 LGS of the ELT. They support all the optical components of the LGSS, as proposed by the optical design. They also support most of the mechanism and devices of the LGSS. The current design is to position all the components in contact with the bench on three pads that defines a plane. The three DOFs out of plane are adjusted with a set of shims that are inserted between the bench and these three pads. The remaining three DOFs in plane are adjusted with shims that are inserted between the baseplate of component and three mechanical stops put on the bench. When a given component is in position, it is screwed by 3 screws passing through the three pads defining the plane. A Spherolinder system permit a precise repositioning.

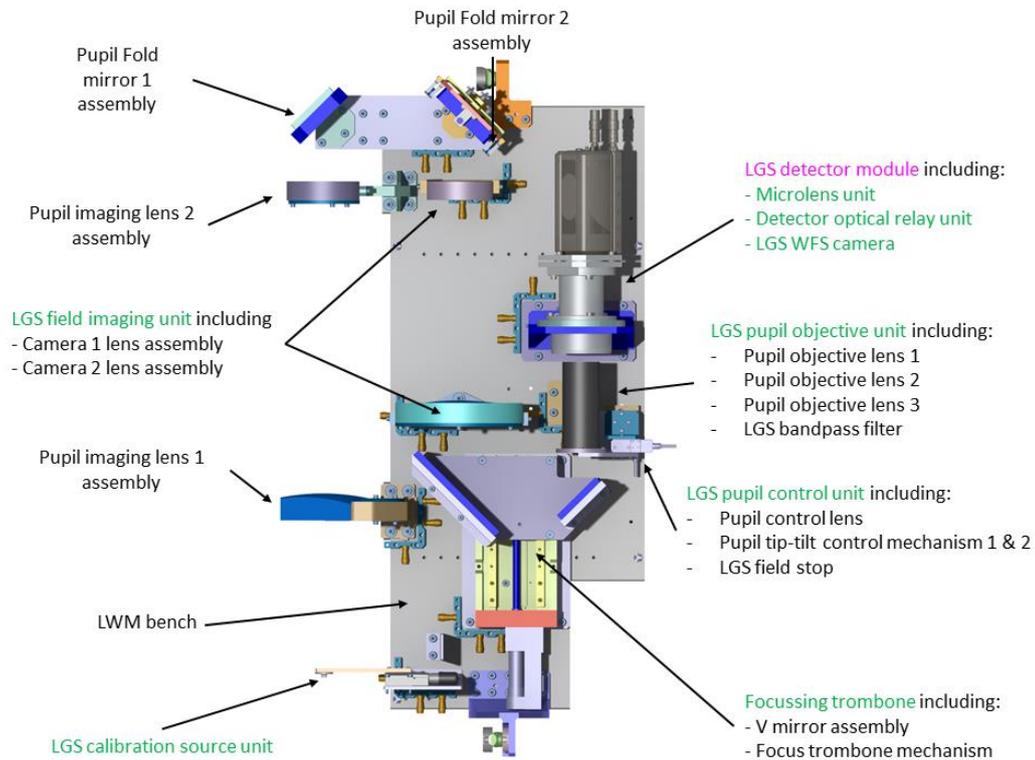

Figure 15: the optical bench assembly

# 4. LGSS MAIT STRATEGY

The LGSS will be assembled and tested under the LAM responsibility. One important point of the LGSS verification is that it will be done on non-elongated source due to constraints on the AIT tool design. In addition, all LGSS requirements exclude the spot elongation effect. The impact of the elongation of the source is analysed by simulation and test will be done on the prototype activities [10] to verify some phenomenon or analysis hypothesis. An important feature of LGSS, as for all AO systems for the ELT, is that it will only meet the deformable mirror at the telescope. This status makes the AIT very different from classical AO system for which complete validation performed in Europe allow go-no go decisions for shipment to Chile. Instead, the complete AO loop (WFS, DM, RTC working together at full speed) will not be assembled before being at telescope and can only be fully tested on-sky. In this context, the LGSS must be treated as a collection of six wavefront sensors that must be tested and validated individually. The LGSS is an assembly of several optical components and the AO performance will be driven by the overall LGS path aberrations. Hence, the overall optical path should be characterized in terms of optical aberrations and pupil distortions. This must be done in representative configurations in which the LGSS will be used, including different point source altitudes at several elevation angles from 60 degrees to Zenith. One of the main goal of the AIT will be to characterize the path aberrations and distortions and produce a look-up table for the LGSS operation at the telescope.

On top of optical aberrations, the LGSS ensures a set of functionalities that must be tested during AIT. In particular:

**Altitude focus Unit (aka "trombone"):** The trombone will be driven both in open-loop according to a LUT and with offsets to this control from LGS signals. The Open-Loop LUT should be built based on an absolute LGS source altitude, during the AIT.

**Pupil mirror:** The pupil mirror will be driven in close loop, hence no need for LUT construction. We will demonstrate the pupil control strategy during AIT as part of the calibration plan.

**Pupil distortion** performance must be validated and characterized. This, again, must be demonstrated at several elevations from 60 degrees to Zenith.

**WFS cameras**: Specific requirements on the cameras and the AIT will demonstrate that those requirements are met.

Finally one main goal of the AIT for LGSS will be to demonstrate operation and calibration templates of LGSS, and to validate them.

Considering all these points, we propose the following strategy for LGSS AIT: we will test the LGSS system in different AIT steps that will correspond to different configurations of the LGSS system.

The first step of the AIT consists on the test of the individual LGSS WFS bench and the individual LGSS detector module. This first step is focussed on the assembly and alignment of six LGSS WFS benches and the validation of the optical performance of each bench. In addition to the verification of the component individual functionality (motor, camera, etc), we measure the optical quality (WFE, pupil distortion, etc) and calibrate the look-up table for reference slopes..

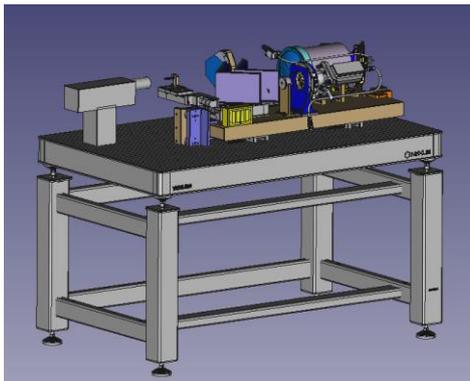

Figure 16: Illustration of the LWM ait – LGSS-1 step.

The second step of the AIT consists mainly on assembly and alignment of LGSS system. It will take place in LAM integration hall and the goal will be to align the WFS bench on the LGSS core structure and on the rotator. In parallel LGSS FM will be assembled to its mechanical structure. Basic commanding test of the whole system will be performed, and first check of operation of LGSS as a whole (motor and camera functionalities).

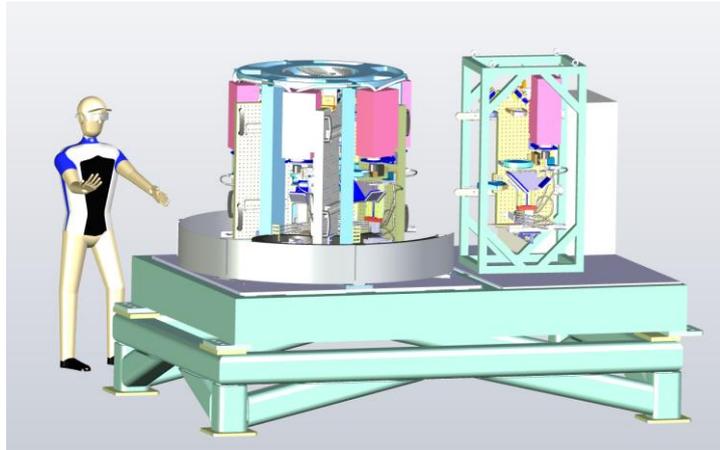
Figure 17: LGSS configuration during LGSS Step 3 ait

The third step of LGSS AIT takes place once LGSS is integrated inside HARMONI top end. This step consists in aligning LGSS onto LGSS dichroic, part of FPRS, on top end, and then to perform LGSS performance test. For LGSS performance test, it is required to develop an ELT telescope simulator with 6 laser guide stars. Studies of the possible design have shown that this ELT telescope simulator will be a quite big instrument, which can also be used to test LTAO performance. After discussion with the project, it has been decided to perform LGSS performance test directly on HARMONI top end, and not during LGSS stand-alone phase, as it was more efficient and useful to develop only one ELT telescope simulator for the project. The goal of the LGSS final tests are:

- to align LGSS on HARMONI
- to test LGSS as the system in HARMONI software configuration (motor functionality, etc)
- to verify the algorithms of LGSS tomography (mis-registration, pupil tracking, reference slopes acquisition, etc)
- to validate the LGSS templates
- to do the performance test linked to AO verification.

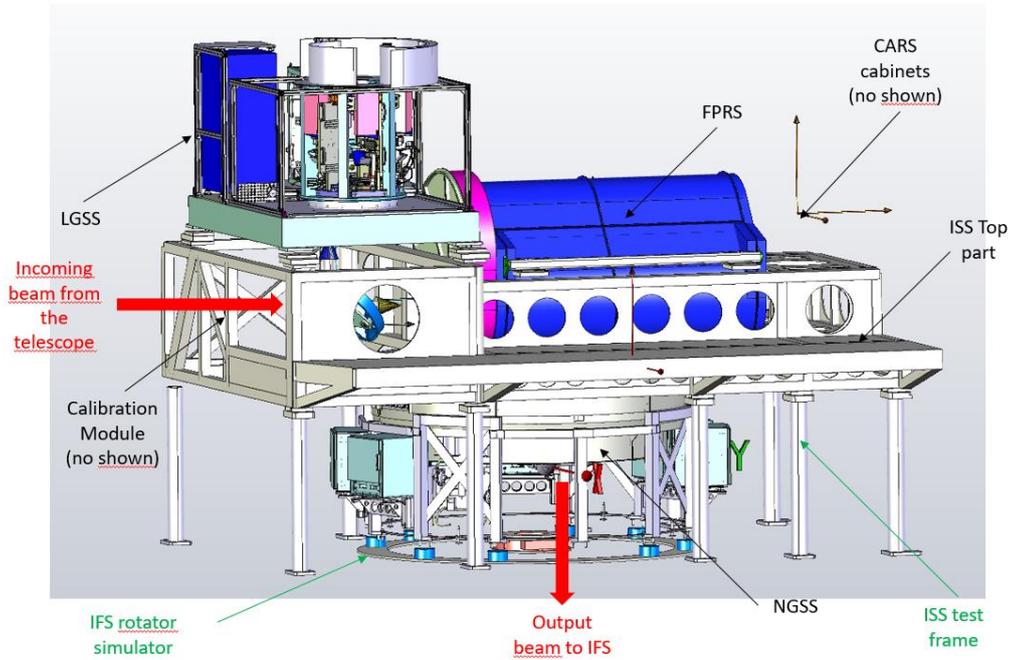

Figure 18: Picture of Top End for Harmoni. LGSS is at the top of Top End close to the entrance of Harmoni.

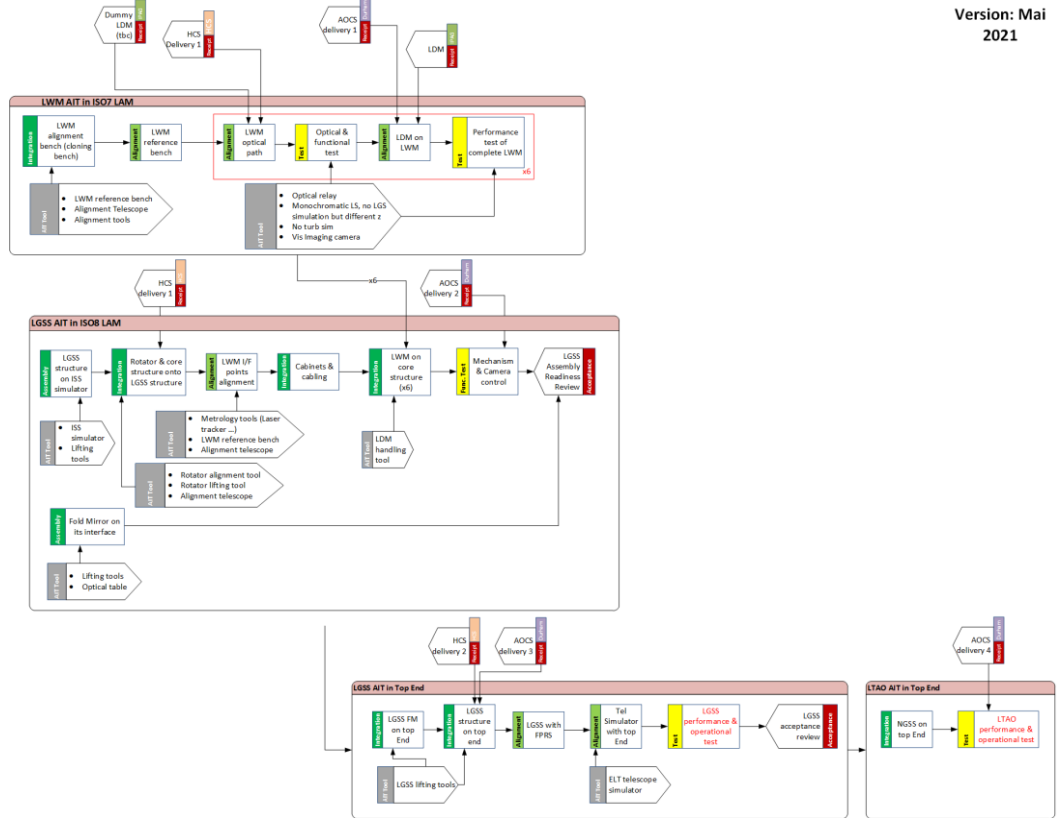

Figure 19: LGSS AIT flow chart

## 5. CONCLUSION

We have described the opto-mechanical designs of the Laser Guide Star Sensors for HARMONI. This system holds the six wavefront sensors used to sense the light from the LGS of the ELT. The description given in this paper are representative of the design done for the Final Design Review that is on going, with a review date around June 2023. If everything goes well, there will be 5 years of AIT of LGSS in stand alone and then in HARMONI in order to be ready for installation in Chile in 2028.

## ACKNOWLEDGEMENTS

HARMONI is an instrument designed and built by a consortium of British, French and Spanish institutes in collaboration with ESO. This work benefited from the support of the French National Research Agency (ANR) from the Programme Investissement Avenir F-CELT (ANR-21-ESRE-0008) and the Action Spécifique Haute Résolution Angulaire (ASHRA) of CNRS/INSU co-funded by CNES.